\documentclass[article,twocolumn,floatfix]{revtex4-1}
\usepackage{epsfig,subfigure,amssymb,amscd,amsmath,graphicx,amsfonts}
\usepackage{times}

\newcommand{\bea}{\begin{eqnarray}}
\newcommand{\eea}{\end{eqnarray}}
\newcommand{\be}{\begin{equation}}
\newcommand{\ee}{\end{equation}}
\newcommand{\ba}{\begin{align}}
\newcommand{\ea}{\end{align}}

\usepackage[usenames]{color}

\begin{document}

\title{Near-zero-energy end states in  topologically trivial spin-orbit coupled superconducting nanowires 
with a smooth confinement}

\author{G. Kells, D. Meidan, and P. W. Brouwer}

\affiliation{Dahlem Center for Complex Quantum Systems and Fachbereich Physik, 
Freie Universit\"{a}t Berlin, Arnimallee 14, 14195 Berlin, Germany}

\begin{abstract}
A one-dimensional spin-orbit coupled nanowire 
with proximity-induced pairing from a nearby $s$-wave superconductor may be in a topological
nontrivial state, in which it has a zero energy Majorana bound state at each end. 
We find that the topological trivial phase may have fermionic end states 
with an exponentially small energy, if the confinement potential
at the wire's ends is smooth.  The possible
existence of such near-zero energy levels implies that the mere observation of a zero-bias peak in the tunneling conductance 
 is not an exclusive signature of a topological superconducting phase even in the ideal clean single channel limit. 
\end{abstract}

\pacs{74.78.Na  74.20.Rp  03.67.Lx  73.63.Nm}

\date{\today} \maketitle

In one dimension, topological superconducting wires are predicted to support a localized Majorana bound state at each end \cite{Kitaev2001}. These Majorana states are particle-hole symmetric and have exactly zero excitation energy. Within the associated degenerate subspace, braiding and exchange operations can be shown to be non-Abelian \cite{Moore91,Read2000,Ivanov2001}, making them potentially useful in a topological quantum computation schemes \cite{Kitaev2003,Kitaev2006,Freedman1998,Nayak2008}. 
It has been recently noted, that in the right parameter regime,  spin-orbit coupled semiconductor 
nanowires with proximity induced superconductivity, should exhibit the required topological 
superconductivity for Majorana pair formation \cite{Lutchyn2010,Oreg2010}. 

In light of these proposals, the experimental observations of zero-bias peaks in normal-metal superconductor tunnel junctions, which are unaffected by small variations of the magnetic field or gate voltages, may indicate the presence of topological superconductivity \cite{Mourik2012,Das2012}.  While these observations are a necessary indicator of the predicted mid-gap Majorana states \cite{Law2009,Flensberg2010}, it is crucial that alternative mechanisms for the zero-bias conductance be ruled out in order for them to be decisive.  One example of such an alternative mechanism applies to quasi-one
dimensional wires with multiple conducting channels \cite{Wimmer2010,Potter2010,Lutchyn2011,Potter2011,Stanescu2011}, for which low-energy fermionic bound states are predicted to appear in the topological as well as in the non-topological phase if the Zeeman energy
exceeds the splitting between transverse subbands \cite{Kells2012,Potter2012,Gibertini2012,Tewari2012,Rieder2012}. Other 
alternative mechanisms involve disorder \cite{Liu2012,Pikulin2012}, possibly in combination with a gapless region at the wire's end \cite{Bagrets2012}. 
These latter findings suggest that clean single-channel wires offer a favorable setting to discern the presence of Majorana end states. Indeed, experiments are progressively approaching this ideal scenario \cite{Mourik2012,Das2012}.

In this letter we show that the original proposals \cite{Lutchyn2010,Oreg2010}
for topological superconductivity in clean one-dimensional semiconductor wires also allow for near-zero-energy end states deep in the topologically trivial phase, provided the potential that 
confines the electrons at the wire's end is smooth. The low energy is a systematic property of these
states, that persists as long as the confining potential and the induced superconductivity are 
smooth functions of position. The existence of such low-energy Andreev states leads to a low-energy
peak in the tunneling conductance in the topologically trivial phase. Since gate-induced confinement potentials 
are typically smooth, the mechanism we describe here may be relevant for the recent 
experiments \cite{Mourik2012,Das2012}. Our analysis is consistent with and explains the observation
of zero-bias conductance peaks in recent numerical simulations of clean semiconductor wires by Prada {\em
et al.} \cite{Prada2012}.

Following the original theoretical proposals \cite{Lutchyn2010,Oreg2010}, we consider a
one-dimensional semiconductor with Rashba spin-orbit 
coupling of strength $\alpha$, subject to a magnetic field with Zeeman energy $B > 0$ and proximity-coupled
to a standard $s$-wave spin-singlet superconductor. Such a system is described by the four-component
Bogoliubov-de Gennes Hamiltonian
\begin{equation}
  H = \left( \frac{p^2}{2m}+V(x)-\mu -B\sigma_x+\alpha p\sigma_y \right) \tau_z + \Delta \sigma_y \tau_x,
  \label{H}
\end{equation}
where $\sigma_{x,y,z}$ and $\tau_{x,y,z}$ are Pauli matrices acting on the spin and particle-hole
degrees of freedom, respectively. Further, $m$ is the effective electron mass, $\mu = p_{\rm F}^2/2m$ 
the chemical
potential, $\Delta$ the proximity induced superconducting gap in the absence of the magnetic field,
and $V(x)$ is the potential that describes the confinement of electrons near the wire's end. 

As shown in Refs.\ \onlinecite{Lutchyn2010,Oreg2010}, the Hamiltonian (\ref{H}) is in a topological phase
with Majorana fermions at its ends if $B > B_{\rm c} = \sqrt{\mu^2 + \Delta^2}$. Here we consider
the topologically trivial regime with weak induced superconductivity, $\Delta \ll B \ll \mu$. The 
condition $B \gg \Delta$ rules out 
spin singlet $s$-wave pairing, so that the induced superconductivity must be of $p$-wave type. However,
unlike in the topological regime, where the model (\ref{H}) effectively admits $p$-wave superconductivity
for one spin channel only, in the non-topological regime $B \ll \mu$ both spin channels acquire 
superconducting correlations. If $B \gg \Delta$ the two spin channels exist as effectively independent 
$p$-wave superconductors in the wire's bulk, but they are coupled at the wire's ends, which gaps out the 
pair of Majorana-like excitations that would have existed at the wire's end for uncoupled channels. 
As we show below,
this coupling is strong if the wire's end is abrupt, but weak if the confinement is smooth, which 
explains the appearance of an Andreev bound state at an energy far below the bulk excitation gap. The 
crucial difference between an abrupt ending and a smooth confinement is that the spin-orbit energy 
$\varepsilon_{\rm so} = \alpha p$ remains finite up to the turning point for a hard-wall confinement, 
whereas $\varepsilon_{\rm so}$ goes to zero continously for a smooth confinement.

In order to arrive at an approximate analytical solution of this problem, we assume that the energies
$B$, $\varepsilon_{\rm so}$, and $\Delta$, are much smaller than the kinetic energy $\mu - V(x)$.
This separation of energy scales breaks down near the turning point at the wire's end, where the 
velocity
\begin{equation}
  v(x) = \sqrt{2 [\mu-V(x)]/m}
\end{equation}
goes to zero. We circumvent this difficulty by solving a modified version of the problem, in which 
the wire has a hard-wall confinement with $V=0$ inside the wire, and a position-dependent spin-orbit 
strength $\tilde \alpha$ with
\begin{equation}
  \varepsilon_{\rm so}(x) = m \alpha v(x) = \tilde \alpha (\tilde x) p_{\rm F}
  \label{eq:epssoalpha}
\end{equation}
to account for the position-dependence of $\varepsilon_{\rm so}$. The two descriptions are essentially
equivalent if the coordinate $\tilde x$ in the hard-wall model is related to the original coordinate
$x$ as
\begin{eqnarray}
\tilde{x} = \int_{x_0}^x dx' \frac{v_{\rm F}}{v(x)},
  \label{eq:tildexx}
\end{eqnarray}
where $v_{\rm F} = \sqrt{2 \mu/m}$ is the Fermi velocity in the hard-wall model. 
Taken together, the relations (\ref{eq:epssoalpha}) and (\ref{eq:tildexx}) ensure that the electrons ``see'' the same Zeeman energy $B$ and spin-orbit energy $\varepsilon_{\rm so}$ as a function of time when they reverse their direction at the wire's end.

The inequalities $B$, $\varepsilon_{\rm so}$, $\Delta \ll \mu$ allow us to linearize the kinetic energy, 
writing
\begin{equation}
  \psi(\tilde{x}) = \psi_{+}(\tilde{x}) e^{i p_{\rm F} \tilde{x}/\hbar} 
  + \psi_{-}(\tilde{x}) e^{-i p_{\rm F} \tilde{x}/\hbar},
\end{equation}
where 
the functions $\psi_{\pm}$ are slow functions of position on the scale $\hbar/p_{\rm F}$.
The function $\psi_{+}$ describes right-moving electrons and left-moving holes, while $\psi_{-}$
describes left-moving electrons and right-moving holes. They are subject to the four-component
Bogoliubov-de Gennes Hamiltonian 
\begin{eqnarray}
  \label{H_pm}
  \tilde{H}_\pm &=&
  \mp i \hbar v_{\rm F} \tau_z\partial_{\tilde{x}} - B \sigma_x\tau_z \pm \tilde{\alpha}(\tilde{x}) 
  p_{\rm F} \sigma_y\tau_z+  \Delta \sigma_y\tau_x,
\end{eqnarray}
and the boundary condition $\psi_{+}(0) = e^{2 i \eta} \psi_{-}(0)$, $\eta$ being a phase shift 
characteristic of the detailed boundary conditions at the wire's end at $\tilde x = 0$.

The normal part of the Hamiltonian (\ref{H_pm}) can be diagonalized by a rotation in spin space. Defining the 
angle $\theta(\tilde{x})$ and the wavenumber $\tilde{k}_{\rm m}(\tilde{x}) > 0$ as
\begin{eqnarray}
  B &=& \hbar v_{\rm F} \tilde{k}_{\rm m}(\tilde{x}) \cos \theta(\tilde{x}),\nonumber \\
  \varepsilon_{\rm so}(\tilde{x}) &=& \hbar v_{\rm F} \tilde{k}_{\rm m}(\tilde{x}) \sin \theta(\tilde{x}),
\end{eqnarray}
a basis change maps the Bogoliubov-de Gennes Hamiltonian (\ref{H_pm}) to $\tilde{H}_{0,\pm} + 
\tilde{H}_{1,\pm}$, with
\begin{eqnarray}
  \tilde{H}_{0,\pm} &=&
  \hbar v_{\rm F} (\mp i \partial_{\tilde{x}} - \tilde{k}_{\rm m} \sigma_z) \tau_z 
  \pm \Delta \sigma_z \tau_x \sin \theta(\tilde{x}) , 
  \label{eq:H0}
  \\
  \tilde{H}_{1,\pm} &=&
  \Delta \sigma_y \tau_x \cos \theta(\tilde{x}) + \frac{\hbar v_{\rm F}}{2} 
  \frac{\partial \theta}{\partial \tilde{x}} \sigma_x \tau_x.
  \label{eq:H1}
\end{eqnarray}
The superconducting pairing in $\tilde{H}_{0,\pm}$ is of $p$-wave type and pairs electrons of
equal spin (in the rotated frame) with $p$-wave gap
\begin{equation}
  \Delta_p = \Delta \sin \theta = 
  \frac{\varepsilon_{\rm so} \Delta}{\sqrt{B^2 + \varepsilon_{\rm so}^2}},
\end{equation}
whereas the superconducting pairing in $\tilde{H}_{1,\pm}$ is of
$s$-wave type and connects electrons of opposite spin.

\begin{figure}
\centering
\includegraphics[width=.4\textwidth,height=0.3\textwidth]{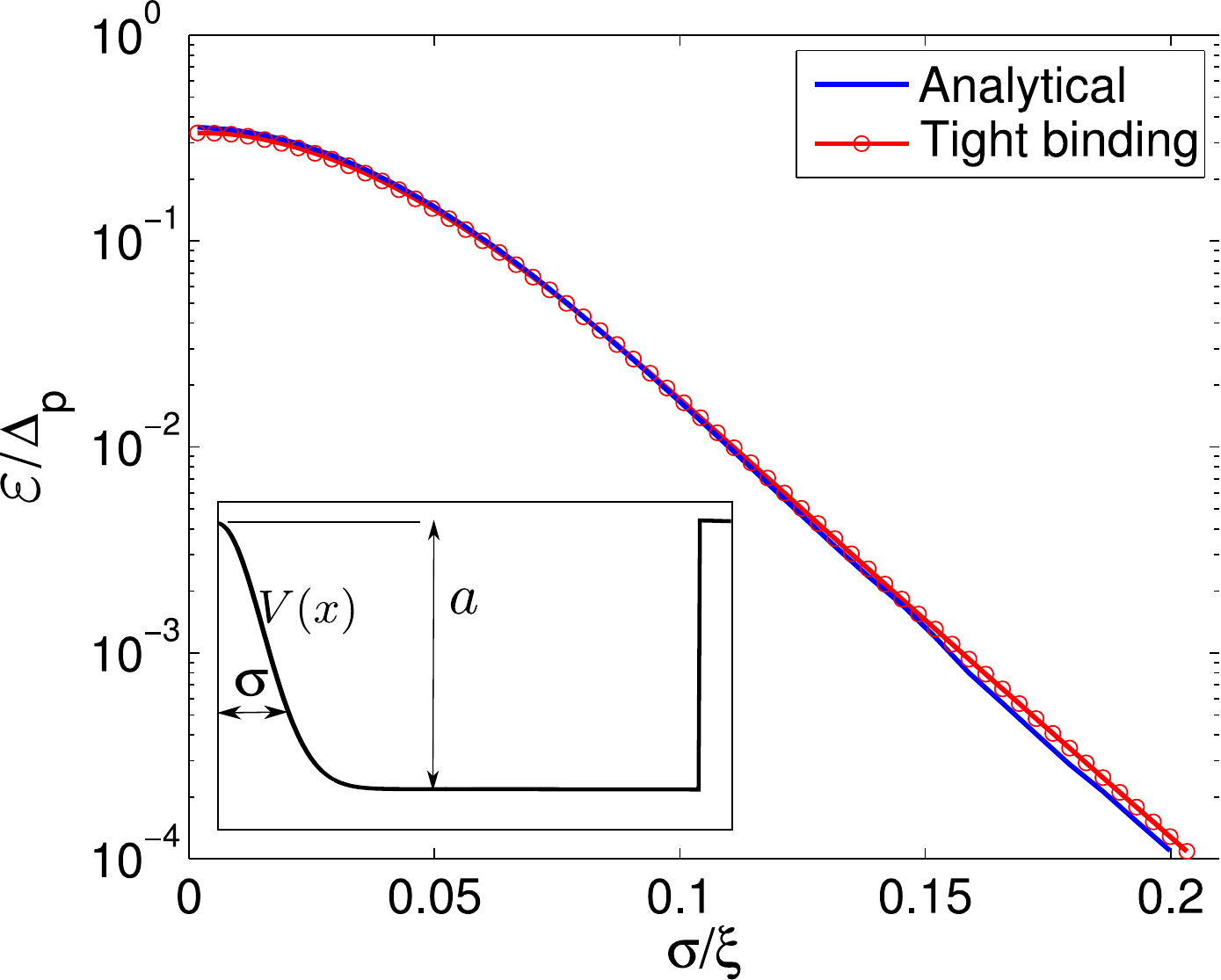}
\caption[]{(Color online) Andreev end-state energy $\varepsilon$, normalized to the bulk excitation gap $\Delta_p$, as a function of adiabaticity parameter $\sigma$.  
The red squares are obtained from a numerical calculation for which the wire is terminated by a smooth potential $V(x)$ 
of the form (\ref{eq:Vconf}) with $a=5 \mu$. Other parameters used in this 
numerical calculation are: $\varepsilon_{\rm so} = 0.1 \mu$, $\Delta = 0.05 \mu$, $B = 0.275 \mu$. The solid blue curve corresponds to Eq.\ (\ref{eq:epsilon}) of the main text.}
\label{fig:sigma}
\end{figure}

In the limit $B \gg \Delta$ and for a smooth confining potential, $\tilde{H}_{1,\pm}$ can be 
treated in perturbation theory. The unperturbed Hamiltonian $\tilde{H}_{0,\pm}$ admits two 
zero-energy end states of Majorana type,
\begin{eqnarray}
  \psi_{\uparrow,\pm} &=&
  \frac{e^{\pm i \eta}}{\sqrt{\tilde{\Omega}}} 
  \left( \begin{array}{c} e^{-i \pi/4} \\ 0 \\ e^{i \pi/4} \\ 0 \end{array} \right)
  e^{\int_0^{\tilde{x}} dx' [\pm i \tilde{k}_{\rm m}(x')-1/\tilde{\xi}(x')]}, \nonumber \\
  \psi_{\downarrow,\pm} &=&
  \frac{e^{\pm i \eta}}{\sqrt{\tilde{\Omega}}} 
  \left( \begin{array}{c} 0 \\ e^{i \pi/4} \\ 0 \\ e^{-i \pi/4} \end{array} \right)
  e^{\int_0^{\tilde{x}} dx' [\mp i \tilde{k}_{\rm m}(x')-1/\tilde{\xi}(x')]}, ~~~~
\end{eqnarray}
where the superconducting coherence length $\tilde{\xi}$ is defined as $\hbar v_{\rm F}/\tilde{\xi}(\tilde{x}) = 
\Delta |\sin \theta(\tilde{x})|$ and $\tilde \Omega$ is a normalization constant. Calculating
the matrix element of $\tilde{H}_{1,\pm}$ between these states, we find that the wire's end
harbors a single Andreev end state with energy
\begin{eqnarray}
  \varepsilon &=&
  \frac{2}{\tilde{\Omega}}
  \left| \vphantom{\int_0^{M^M_M}}
  \int_0^{\infty} d \tilde{x}
  \left[ 2 \Delta \cos \theta(\tilde{x}) 
  + \hbar v_{\rm F} \frac{\partial \theta}{\partial \tilde{x}} \right] 
  \right. \nonumber \\ && \left. \mbox{} \times
  \cos \left[ 2 \int_0^{\tilde{x}} dx'
  \tilde{k}_{\rm m}(x') \right]
  e^{-2 \int_0^{\tilde{x}} dx' 1/\tilde{\xi}(x')} 
  \right|.
\end{eqnarray}
Returning to the parameters of the original model (\ref{H}), the energy $\varepsilon$ of the Andreev end 
state reads
\begin{eqnarray}
  \varepsilon &=&
  \frac{2 B}{\Omega}
  \left|
  \int_{x_0}^{\infty} dx
  \frac{2 \Delta \sqrt{B^2 + \varepsilon_{\rm so}(x)^2} - \hbar \alpha (dV/dx)}
  {v(x) [B^2 + \varepsilon_{\rm so}(x)^2]}
  \right. \nonumber \\ && \left. \mbox{} \times
  \cos \left[2 \int_{x_0}^{x} dx'
  k_{\rm m}(x') \right]
  e^{-2 \int_{x_0}^{x} dx' 1/\xi(x')} 
  \right|,
  \label{eq:epsilon}
\end{eqnarray}
where $\hbar v(x) k_{\rm m}(x) = m \alpha \xi(x) \Delta/\hbar = \sqrt{B^2 + \varepsilon_{\rm so}^2}$ and
\begin{equation}\label{eq:norm}
  \Omega = 4 \int_{x_0}^{\infty} dx \frac{e^{-2 \int_{x_0}^{x} dx' 1/\xi(x')}}{v(x)}.
\end{equation} 
For the simple example that $V(x)$ has a linear dependence on $x$ near the wire's end, $V(x) = \mu - V' x$, 
with the condition that $V' \ll B \Delta/\hbar \alpha$ --- which ensures that that 
$\varepsilon_{\rm so} \ll B$ throughout
the entire integration range ---, a closed-form expression can be obtained and one finds 
\begin{equation}\label{linear_E}
  \varepsilon \approx \Delta
  e^{-B^3/(\hbar \alpha V' \Delta)}.
\end{equation} 
This result should be compared with the energy of the Andreev end state for a hard wall, which is
\begin{equation}
  \varepsilon = \left( \frac{\Delta^2}{2 B^2} + \frac{\varepsilon_{\rm so}}{\Delta} \right) \Delta
\end{equation}
if $\varepsilon_{\rm so}$, $\Delta \ll B$. The energy (\ref{linear_E}) is essentially zero --- even if
compared with the $p$-wave gap $\Delta_p$ --- for a range of magnetic field far below the critical 
field $B_{\rm c}$ at which the transition to the topological phase takes place. (Incidentally, even 
with hard-wall boundary conditions, the Andreev end-state energy $\varepsilon$ may be small in
comparison to $\Delta_p$ if $B$ is sufficiently large in comparison to $\varepsilon_{\rm so}$ and 
$\Delta$.) Experimentally, the difference between the finite excitation energy $\varepsilon$ of Eq.\
(\ref{linear_E}) and the strict zero-energy of the Majorana bound states may be difficult to resolve.

Both the splitting from the singlet pairing 
[first term in Eq.\ (\ref{eq:H1})] and the splitting from the non-adiabaticity of the confining potential 
[second term in Eq.\ (\ref{eq:H1})] vanish exponentially in the limit of a smooth, adiabatic confinement, 
provided the Zeeman energy $B$ sufficiently far exceeds $\Delta$. This is the main result of this letter.
We stress that in the weak pairing limit, the splitting $\varepsilon$ decreases with increased coherence
length, and therefore cannot be simply understood as the result of the (small) separation between the
turning points for electrons with opposite spin. (Indeed, no such separation is present in the
effective hard-wall model used for our calculation!)

\begin{figure}
\centering
\includegraphics[width=.4\textwidth,height=0.3\textwidth]{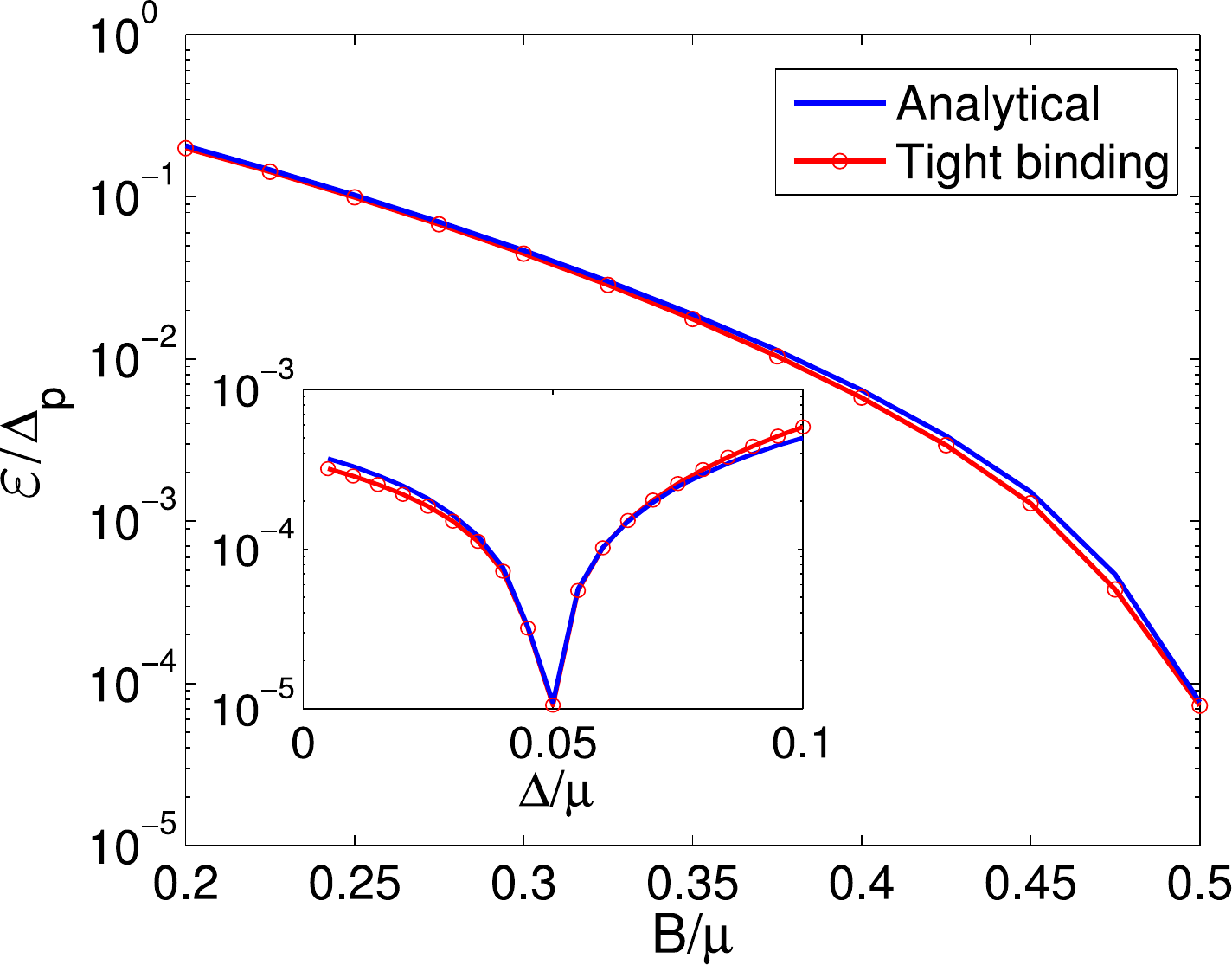}
\label{fig:BS4}
\caption[]{(Color online) Andreev end-state energy $\varepsilon$ as a function of the Zeeman energy $B$ (main panel) and induced superconducting gap parameter $\Delta$ (inset). The parameters of the numerical calculation are: $\varepsilon_{\rm so} = 0.1 \mu$, $\Delta = 0.04 \mu$ 
(main figure), $B = 0.5 \mu$ (inset). The confining potential has the form (\ref{eq:Vconf})
with $a = 5 \mu$ and $\sigma = 1.273 (h/p_{\rm F})$.
}
\label{fig:BDS4}
\end{figure}

We have compared the theoretical predictions to numerical tight-binding simulations of a discretized
version of the Hamiltonian (\ref{H}). For this purpose we choose the confining potential
\be
  V(x) = \left\{ \begin{array}{ll} a e^{-x^2/2 \sigma^2} & x > 0,\\
  \infty & x < 0, \end{array} \right.
  \label{eq:Vconf}
\ee
where the parameter $\sigma$ controls the degree of adiabaticity. Figures \ref{fig:sigma} and
\ref{fig:BDS4} show representative
results of the numerical calculation, together with the analytical prediction of Eq.\ (\ref{eq:epsilon}).

As discussed above, the small energy $\varepsilon$ of the Andreev endstates results from the 
ineffectiveness of a smooth potential to couple the two Majorana modes for the two spin channels. This 
near-degeneracy will be lifted in the presence of perturbations with an abrupt spatial dependence
that couple the different spin-orbit bands. Examples of such perturbations are scattering from 
point-like impurities (which couple left-moving and right-moving particles), or a the abrupt vanishing
of the pairing potential, which happens, {\em e.g.}, if not all of the semiconducting wire is covered
with the superconducting contact. The Andreev end-state energy $\varepsilon$ in the presence of a point impurity 
with potential $U \delta(x-x_{\rm i}) \tau_z$ is (to first order in $U$)
\begin{eqnarray}
  \varepsilon &=& 
  \frac{4 U \alpha m e^{-2 \int_{x_0}^{x_{\rm i}} dx \xi^{-1}(x)} }
  {\Omega \sqrt{B^2 + \varepsilon_{\rm so}(x_{\rm i})^2}}
 \left| \sin \left[ 4 \eta + \int_{x_0}^{x_{\rm i}} dx \frac{m v(x)}{\hbar} \right]\right|.
  ~~~
\end{eqnarray}
For the example of a slowly varying potential $V(x) = \mu - x V'$ with a linear dependence on position, this gives
\begin{eqnarray}
\nonumber
  \varepsilon &=& 2 U m \sqrt{\frac{\alpha^{3} V' \Delta}{\pi \hbar B^3}}e^{-\frac{2m \alpha \Delta}{\hbar B}(x_i-x_0)}\\
&& \mbox{} \times
   \left|\sin \left[ 4 \eta +\frac{2\sqrt{2m V'}}{3\hbar}(x_i-x_0)^{3/2} \right] \right|
 \end{eqnarray}

\begin{figure}
\centering
\includegraphics[width=.4\textwidth,height=0.3\textwidth]{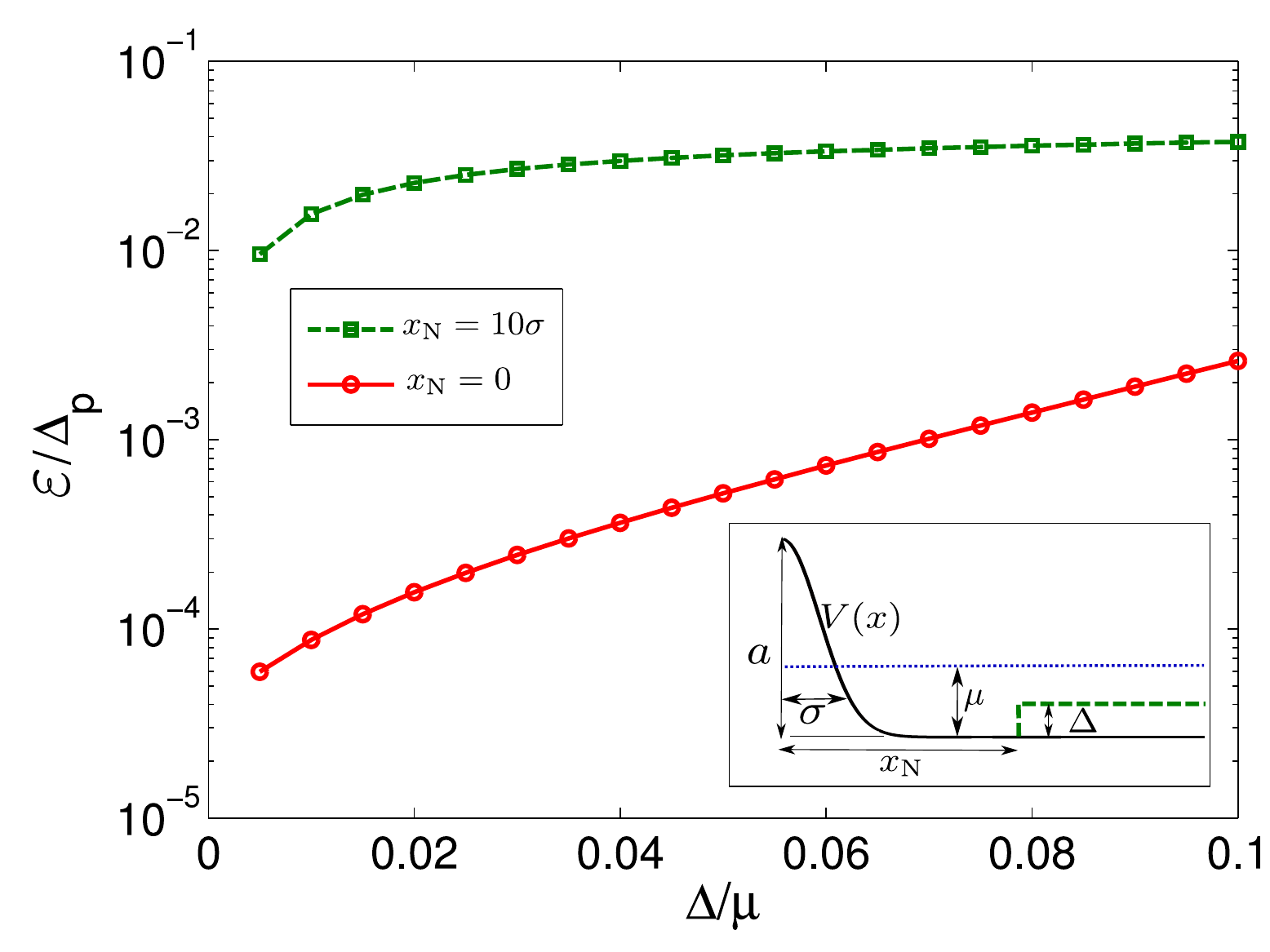}
\caption[]{(Color online) Andreev end-state energy $\varepsilon$ versus the induced superconducting gap parameter $\Delta$ with (squares) and without (circles) an abrupt termination of the order parameter $\Delta$ at position $x_{\rm N}$. The inset shows the functional form of the superconducting order $\Delta(x) = \Delta  \theta(x-x_{\rm N})$. In both cases the wire is terminated by a smooth 
potential $V(x)$ of the form (\ref{eq:Vconf}) with parameters with $a=5 \mu$ and $\sigma \approx 3.183 h/p_{\rm F}$. Other parameters used in this numerical calculation are $\varepsilon_{\rm so} = 0.1 \mu$ and $B = 0.275 \mu$.
}
\label{fig:compare}
\end{figure}

In the case that the order parameter vanishes abruptly, $\Delta(x) = \Delta\Theta(x-x_{\rm N})$ (see inset of Fig.~\ref{fig:compare}), the end-state energy can be obtained directly from Eq.\ (\ref{eq:epsilon}). For the special case that  the entire potential modulation occurs in the normal region,  the discontinuity in $\Delta$  contributes to the end-state energy by the amount 
\begin{eqnarray}
\nonumber
  \varepsilon &=&\frac{2 \hbar B}{\Omega}\frac{1} {m\alpha v_F}
  \mbox{Re}\, \frac{e^{2 i \int_{x_0}^{x_N} dx k_{\rm m}(x)}}
  {1 + i k_{\rm m} \xi},
\end{eqnarray}
where
\begin{eqnarray}
  \Omega =\frac{2\xi}{v_F} + 4 \int_{x_0}^{x_{\rm{N}}} dx \frac{1}{v(x)}
\end{eqnarray}
and $k_{\rm m}$, $\xi$, and $v_{\rm F}$ are the asymptotic values for $x > x_{\rm N}$. Figure \ref{fig:compare} compares numerical simulations of the model (\ref{H}) with and without an 
abrupt change in the superconducting order parameter $\Delta$.

\begin{figure}[h]
\centering
\includegraphics[width=.4\textwidth,height=0.3\textwidth]{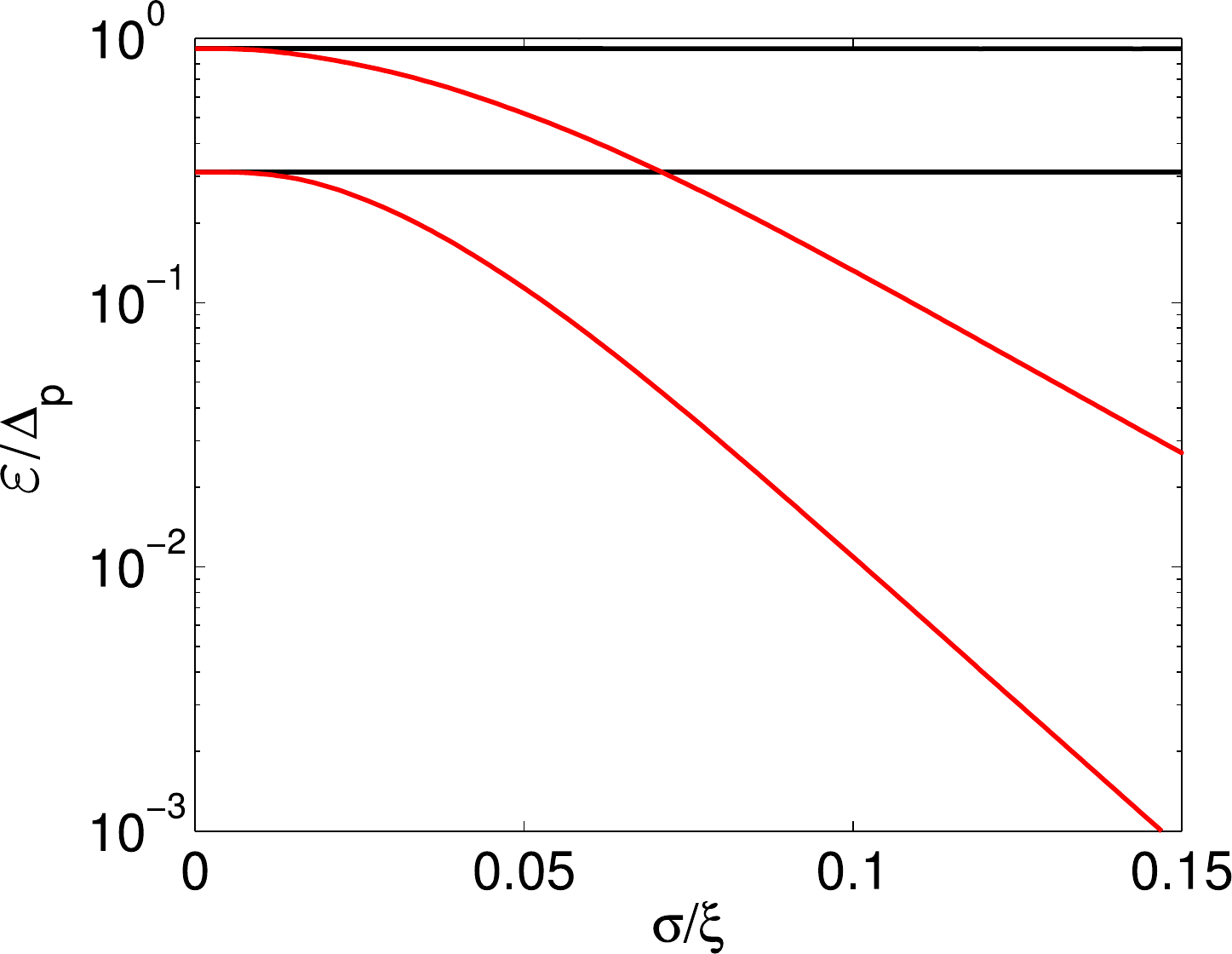}
\caption[]{(Color online) Andreev end-state energies $\varepsilon$ for a two-dimensional wire with two transverse channels, normalized by the bulk excitation gap, as a function of the adiabaticity parameter $\sigma$. The Hamiltonian is given by the two-dimensional extension of Eq.\ (\ref{H}), which has the spin-orbit coupling term $\alpha p_x \sigma_y \tau_z - \alpha p_y \sigma_x$. Parameters in the numerical calculation are $B = 0.1667 \mu$, and wire width $W = 1.225 h/p_{\rm F}=0.054 \xi$.  The spin-orbit energies of the two transverse bands (calculated for $B=0$) are  $\varepsilon_{{\rm so},1} = 0.074 \mu$, $\varepsilon_{{\rm so},2} = 0.042\mu$. 
}
\label{fig:multichannel}
\end{figure}

Up to this point, our discussion has focused on one-dimensional semiconductor wires with a single transverse channel. Our arguments continue to be valid for multichannel wires. In this case, for $B \gtrsim \Delta$  each transverse channel is in a separate effectively spinless $p$-wave superconducting state. Hard-wall boundary conditions at the wire's end couple the channels, which gaps out the end states, up to the possible exception of a single Majorana end state if the total number of channels $N$ (counting spin) is odd. In general, a coupling exists between spin degenerate channels with the same transverse mode, as well as between channels with different transverse modes, although the ``off-diagonal'' coupling is small if the wire width is much smaller than the superconducting coherence length $\xi$ because of an approximate chiral symmetry \cite{Kells2012,Tewari2011}. By the mechanism discussed above, a smooth confinement strongly reduces the diagonal and the off-diagonal couplings between channels, giving rise to $\mbox{int}\, (N/2)$ low-energy fermionic states at the wire's end. This is illustrated in Fig.\ \ref{fig:multichannel} for the case of a semiconducting wire with two spin-degenerate transverse channels.

In conclusion, we have shown that a sufficiently smooth confinement potential at the wire's ends leads
to the existence of low-energy Andreev end states even in the topologically trivial 
phase of one dimensional proximity coupled nanowires. These results could be relevant for recent experiments 
\cite{Mourik2012,Das2012}, in which the confinement at the nanowire ends is gate-induced.  
The presence of such low energy Andreev states would lead to near zero conductance peaks deep in the topologically 
trivial parameter regime. This low energy  peak hinders  the experimental verification of a topological 
superconducting phase via tunneling density of states, even in the ideal single channel case.

We gratefully acknowledge discussions with J. Danon, F.\ Von Oppen, Y.\ Oreg, and F.\ Pientka. This work is supported by the Alexander von Humboldt Foundation in the framework of the Alexander von Humboldt Professorship, endowed by the Federal Ministry of Education and Research.

\end{document}